\documentclass[10pt]{article}

\usepackage{graphicx}
\usepackage{amsmath}
\usepackage{amssymb}
\usepackage{verbatim}
\usepackage{times}

\DeclareGraphicsExtensions{.eps}

\date{} 

\begin{document}

\title{Social Dilemmas and Cooperation in Complex Networks}

\author{Marco Tomassini \and Leslie Luthi \and Enea Pestelacci
\\ \small Information Systems Department,
University of Lausanne, Switzerland}

\maketitle


\begin{abstract}

\noindent In this paper we extend the investigation of cooperation in some classical
evolutionary games on populations were the network of interactions among individuals is
of the scale-free type. We show that the update rule,
the payoff computation and, to some extent the timing of the operations, have a marked influence on the transient dynamics and on the amount of cooperation that can
be established at equilibrium. We also study the dynamical behavior of the populations and
their evolutionary stability.

\end{abstract}

\section{Introduction and Previous Work}

The object of game theory is the analysis of situations where the different social actors
have conflicting requirements and individual decisions will have a mutual influence on each other\cite{myerson91}. In this framework, and
due to their importance as simplified models of many common important socio-economic
 situations, the \textit{Prisoner's
 Dilemma} (PD) and the \textit{Snowdrift} (SD) games have received much attention in the literature. 
According to game theory, the PD and the
SD are paradigmatic examples of games in which cooperative attitude
should vanish in the PD, and should be limited to a given fraction in the 
SD. This is also the case when large populations of individuals play the game
pairwise in a random manner and anonimously, as prescribed by evolutionary game theory \cite{weibull95}.
In spite of this, numerical simulations of the PD have convincingly shown that,
when the population of players possesses a spatial structure,
a certain amount of cooperation can emerge and remain stable. Nowak and May \cite{nowakmay92} were the first to empirically show this using
a population structured as a square lattice where each site is a player. Standard evolutionary game theory is based
on an infinite (or very large) population model, and on the random pairing of two players
at each time step. This amounts to a \textit{mean-field} spatially homogeneous model.
The square grid is also spatially homogeneous but the absence of random mixing enables
the formation of clusters of cooperators, which allows for more frequent positive encounters between
cooperators than would be possible in the mean-field case.
More recently, it has become apparent that many real networks are neither regular nor random
graphs; instead, they have short diameters, like random graphs, but much higher clustering
coefficients than the latter, i.e. they have more local structure. These networks are
collectively called \textit{small-world} networks (see \cite{newman-03} for a recent review).
Many technological, social, and biological networks are now known to be of this kind.
Social networks, in addition, show recognizable community structure \cite{newman-girvan-2004-69,gonzalez06}.
Since evolutionary PD or SD games are metaphors for
conflicting social interactions, the research attention has recently shifted from random
graphs and regular lattices towards better models of social
interaction structures \cite{social-pd-kup-01,karate-pd-03,santos-pach-05,tom-lut-gia-hd-06}.

Recently, Santos and Pacheco \cite{santos-pach-05} presented a numerical
study of the evolution of cooperation on (static) scale-free (SF) networks for the PD and the SD games.
Their main result was that, in contrast with what one observes in mixing populations or on regular lattices,
much higher levels of cooperation are sustainable on this kind of graphs, both for
the PD as well as the SD. These results are
obviously interesting and encouraging for cooperation but they prompt a number of questions.
First of all, Bar\'abasi--Albert  or correlationless configuration SF graphs 
\cite{alb-baraba-02} that were used
in \cite{santos-pach-05} are not faithful representations of most typical social
networks. In fact, although social interaction networks where the degree distribution can be well described by a power-law have been found \cite{liljeros01,hans06}, several recent studies show that social networks in general do not have a pure power-law degree distribution function,
as they often show signs of exponential decay of the tail of the
distribution \cite{am-scala-etc-2000,newman-collab-01-1}. In addition, they usually have more clustering than pure scale-free graphs \cite{newman-03}.  
Nevertheless, model SF networks are a useful bounding case to study as they are closer to
typical social networks than other more artificial kind of graphs, such as Watts--Strogatz
small worlds \cite{watts-strogatz-98}. 
A second aspect of social networks that is not captured by fixed graph structures is that they are not static; rather, the number
of vertices and the links between them continuously evolve as social actors come and go, and
relationships are created or abandoned. Dynamical features such as these have been introduced
in evolutionary games, among others, in
\cite{zimm-et-al-04,zimm-egui-05,biely-05,lut-giac-tom-al-06,santos-plos-06}. However, in this paper we only focus on the static aspects of the interaction networks. 
In other words, we make the hypothesis that the network is at equilibrium and that
network dynamics are either absent, or their time scale is longer
(slower) with respect to the strategy-change dynamics. This proves to be a useful approach, especially for social acquaintance networks. 

In the following we present a brief introduction to the games studied. This is followed
by a discussion of the population model and of individual's payoff calculation scheme for the players in a complex network. Next we describe the
numerical simulations and their results, including a study of evolutionary stability.
We finally present our conclusions.

\section{Two Social Dilemmas}
\label{dilemmas}

Let us first recall a few elementary notions on the PD and the SD.
These are two-person, symmetric games in which each player has two possible strategies:
cooperate (C) or defect (D). In strategic form, also known as normal form, these games have the payoff bi-matrix of table \ref{payofft}.
\begin{table}[htb]
\normalsize
\begin{center}
\begin{tabular}{c|cc}
 & C & D\\
\hline
C & (R,R) & (S,T)\\
D & (T,S) & (P,P)
\end{tabular}
\caption{Payoff matrix for a standard two-person, two-strategies game (see text).\label{payofft}}
\end{center}
\end{table}
In this matrix, R stands for the \textit{reward}
the two players receive if they
both cooperate, P is the \textit{punishment} for bilateral defection, and T  is the
\textit{temptation}, i.e. the payoff that a player receives if it defects, while the
other cooperates. In this latter case, the cooperator gets the \textit{sucker's} payoff S.
For the PD, the payoff values are ordered numerically in the following way: $T > R > P > S$, 
while in the SD game $T > R > S > P$. Defection is always the best rational individual choice in the PD -- (D,D) is the unique Nash equilibrium and also an evolutionary stable strategy (ESS).
Mutual cooperation  would be preferable but it is a strongly dominated strategy. Thus the
dilemma is caused by the ``selfishness'' of the actors.
 
 In the SD, when both players defect they each get the lowest payoff;
(C,D) and (D,C) are Nash equilibria of the game in pure strategies, and there is
a third equilibrium in mixed strategies where strategy D is played
with probability $1/(2\beta-1)$, and strategy C with probability $1 - 1/(2\beta-1)$, where
$\beta$ is another name for the temptation $T$, used in biological circles.
The dilemma in this game is caused by ``greed'', i.e. players have a strong incentive
to ``bully'' their opponent by playing D, which is harmful for both parties if the outcome produced is (D,D).

\section{Numerical Simulations}
\label{simul}

The two games were simulated in \cite{santos-pach-05} on Barab\'asi-Albert (BA) \cite{alb-baraba-02} and configuration model \cite{newman-03} scale-free networks of size $10^4$ over $10^4$ time steps, using a discrete analogue of \textit{replicator dynamics} equations \cite{weibull95,hauer-doeb-2004}. 
The customary rescaling of the payoff values was used such that there is only one independent parameter.
For the PD, setting $R=1$, $P=S=0$, leaves $T=b > 1$ to be the only parameter (temptation). For
the SD, $T$ is set equal to $\beta > 1$, $R = \beta-1/2$, $S= \beta-1$, and $P=0$, which
makes the cost-to-benefit ratio of mutual cooperation $r=1/(2\beta-1)$ the only parameter.
For the sake of comparison, our simulations were done under the same conditions as in \cite{santos-pach-05} ($10^4$ players and $10^4$ time steps).

However, replicator dynamics is not the only possibility for updating the agents' strategies in discrete, finite populations of players using hard-wired strategies. Moreover, in small 
non degree-homogeneous populations, the mathematical requirements behind the replicator dynamics,
strictly speaking, are not satisfied \cite{nowak-et-al-finite-04}. 
Thus, we extended the investigation by simulating  an \textit{imitate the best} evolution rule according to which 
an individual $i$ will adopt the strategy of the player with the highest
payoff among its neighbors and itself. If a tie occurs, the winner is chosen uniformly at random between the best. This rule is deterministic and was the original
rule used in \cite{nowakmay92}.

Concerning the calculation of an individual's payoff, there are several possibilities.
A player's payoff may be defined as the sum (\textit{accumulated payoff}) of all
pair interactions with its nearest neighbors, which is the form used for instance in
\cite{santos-pach-05}. Another possibility consists in using
\textit{average payoff}, which is the accumulated payoff divided by the number of interactions.
Accumulated and average payoff give the same results 
when considering degree-homogenous networks such as lattices. 
Accumulated payoff seems more logical in degree-heterogeneous networks such as scale-free
graphs since it
reflects the very fact that players may have different numbers of neighbors in the network.
Average payoff, on the other hand, smooths out the possible differences although it might
be justified in terms of number of interactions that a player may sustain in a given time.
For instance, an individual with many connections is likely to interact less often with each of its neighbors than another that has a lower number of connections. Also, if there is a cost to maintain a relationship,
average payoff will roughly capture this fact, while it will be hidden if one uses
accumulated payoff.
For the sake of comparing the two extreme views, here we use both accumulated and average payoff.

\begin{figure} [!ht]
\begin{center}
\begin{tabular}{cc}
	\mbox{\includegraphics[width=5.25cm, height=4cm]{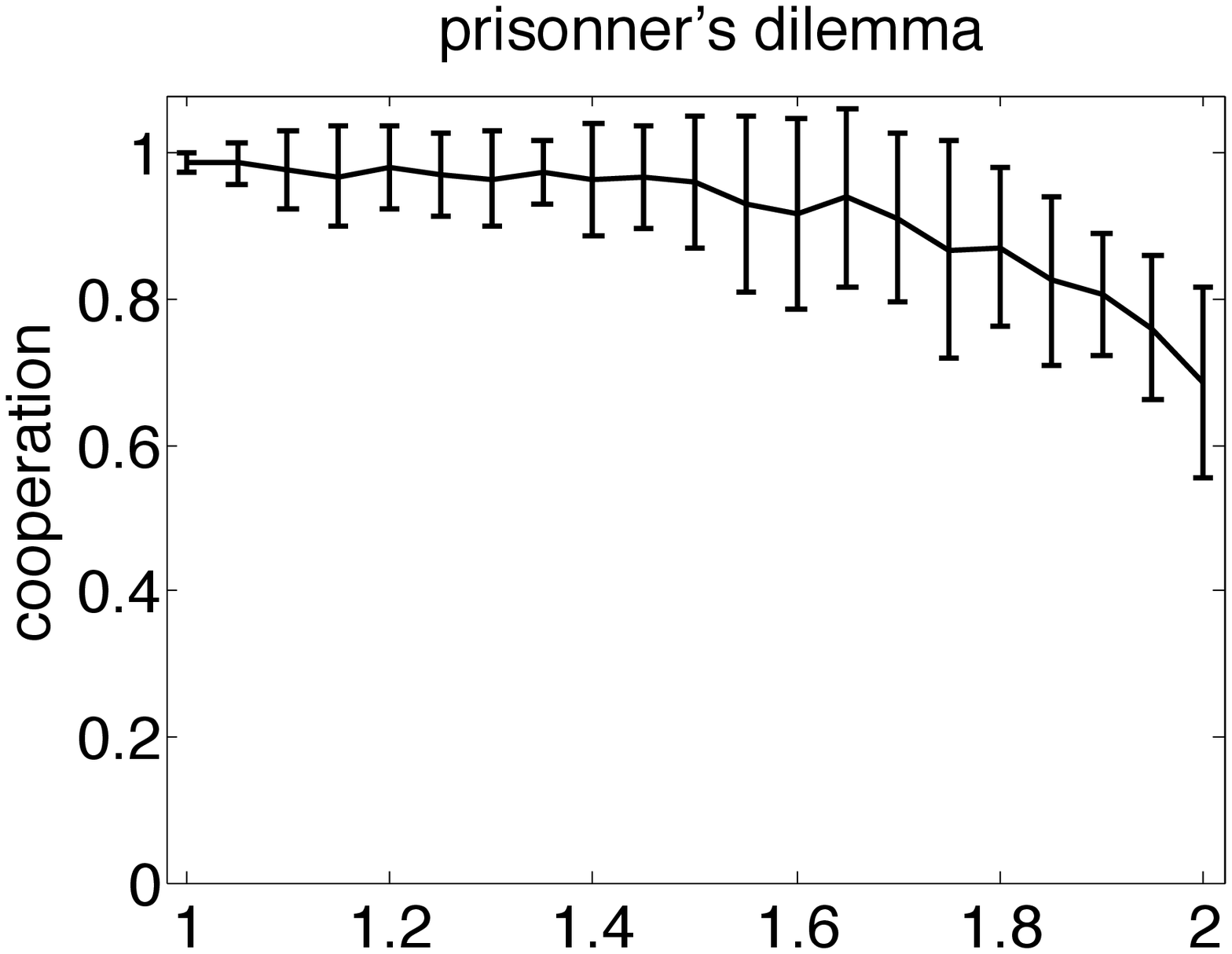}} \protect & 
	\mbox{\includegraphics[width=5cm, height=4.02cm]{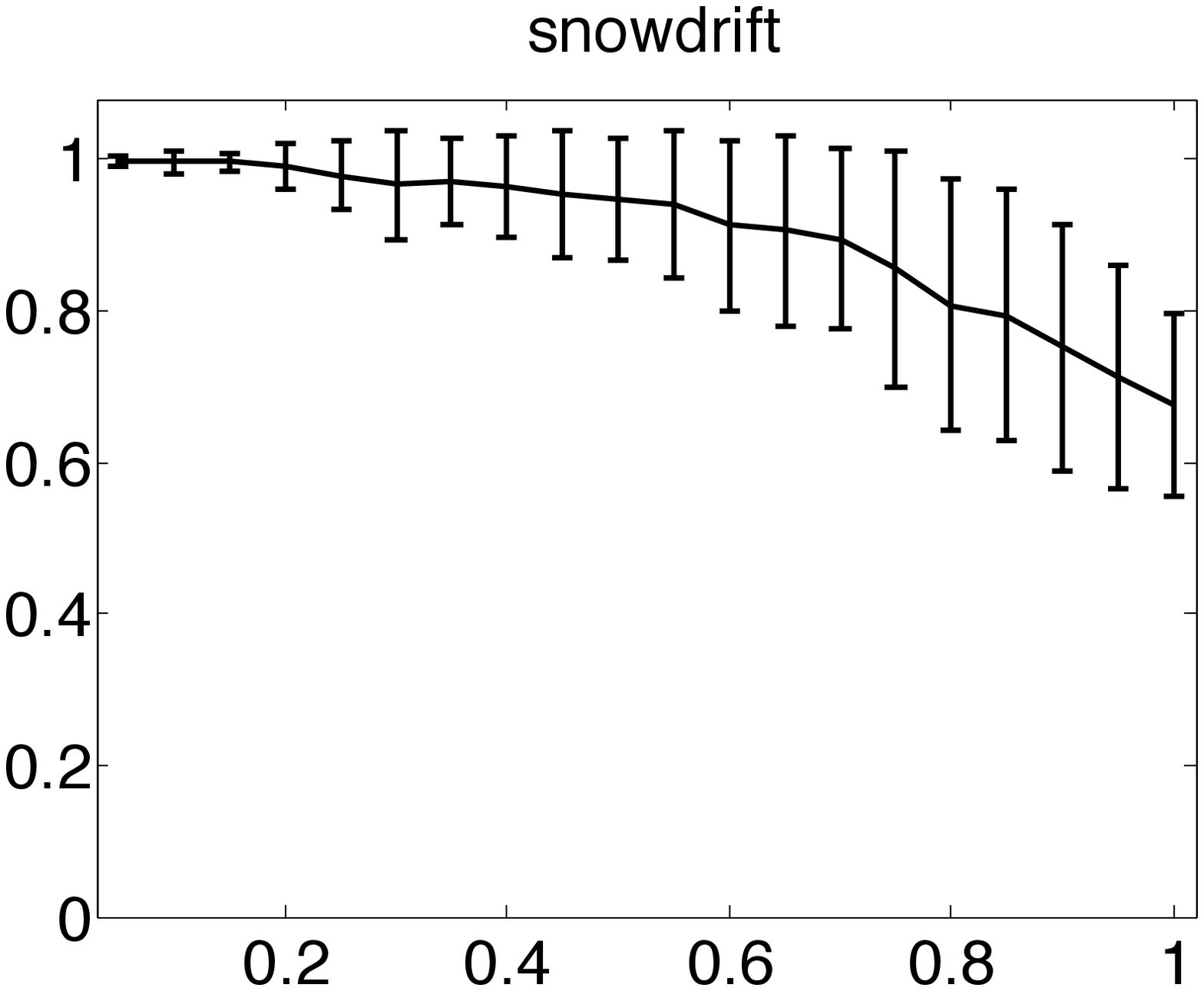}}   \protect \\
	\mbox{\includegraphics[width=5.25cm, height=3.9cm]{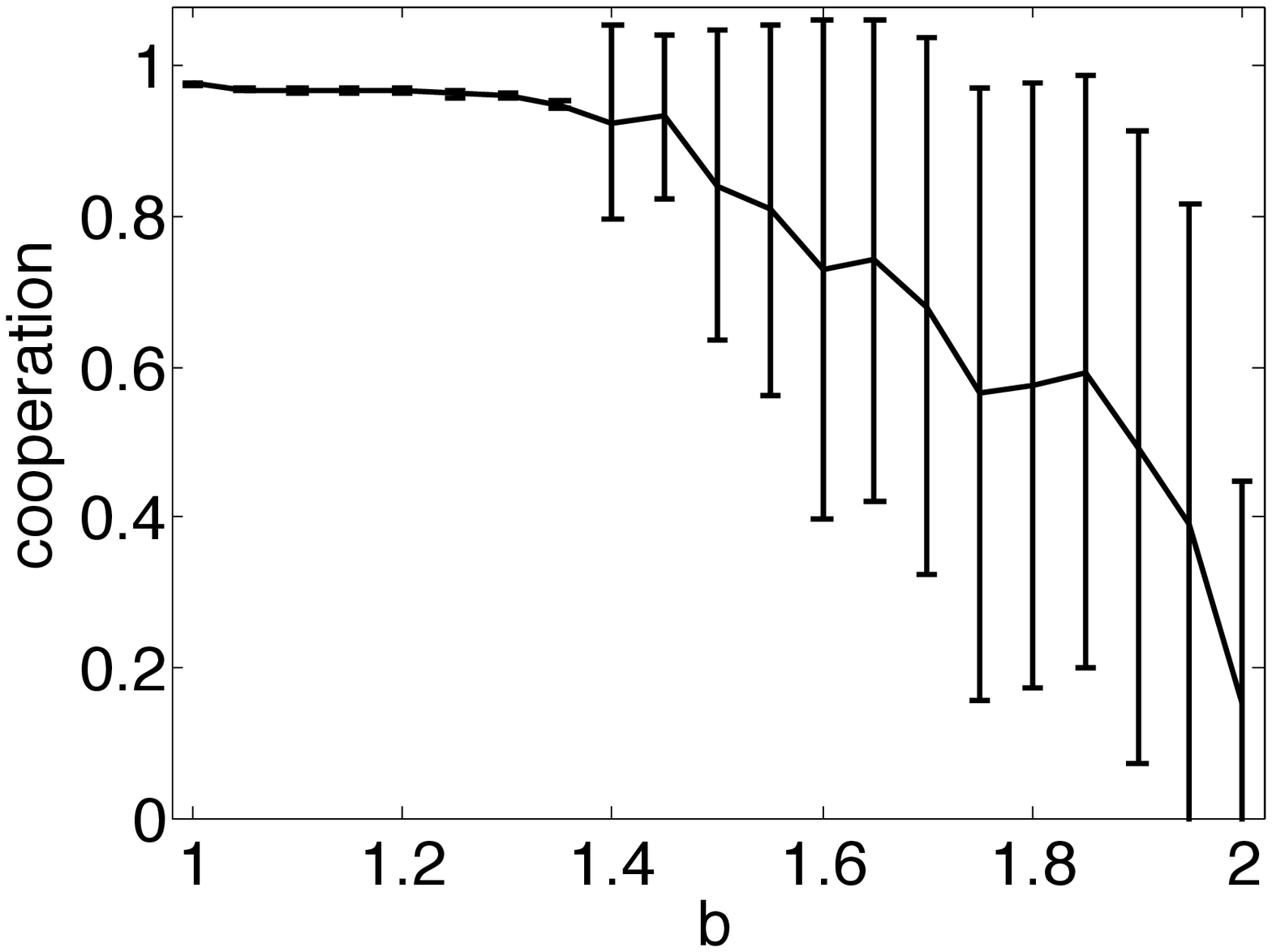}} \protect & 
	\mbox{\includegraphics[width=5cm, height=3.95cm]{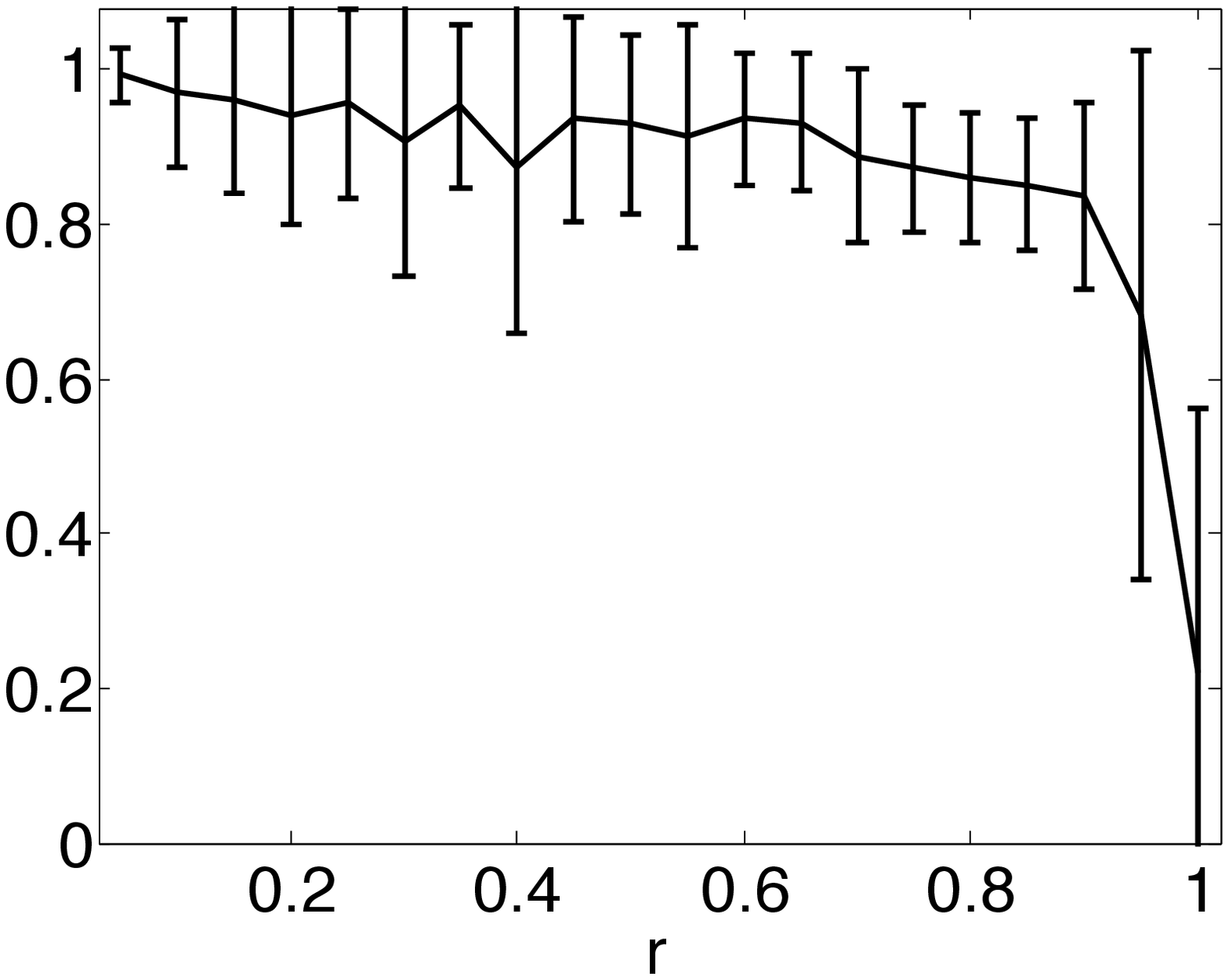}}   \protect \\
\end{tabular}
\caption{Fraction of cooperators on SF BA networks of size $10^4$ and average degree $\bar k=4$ with accumulated payoff and synchronous dynamics. Mean values
over 50 runs. Upper figures: replicator dynamics; lower figures: imitation of the best. \label{acc-payoff}}
\end{center}
\end{figure}

\begin{figure} [!ht]
\begin{center}
\begin{tabular}{cc}
	\mbox{\includegraphics[width=5.5cm, height=4.2cm]{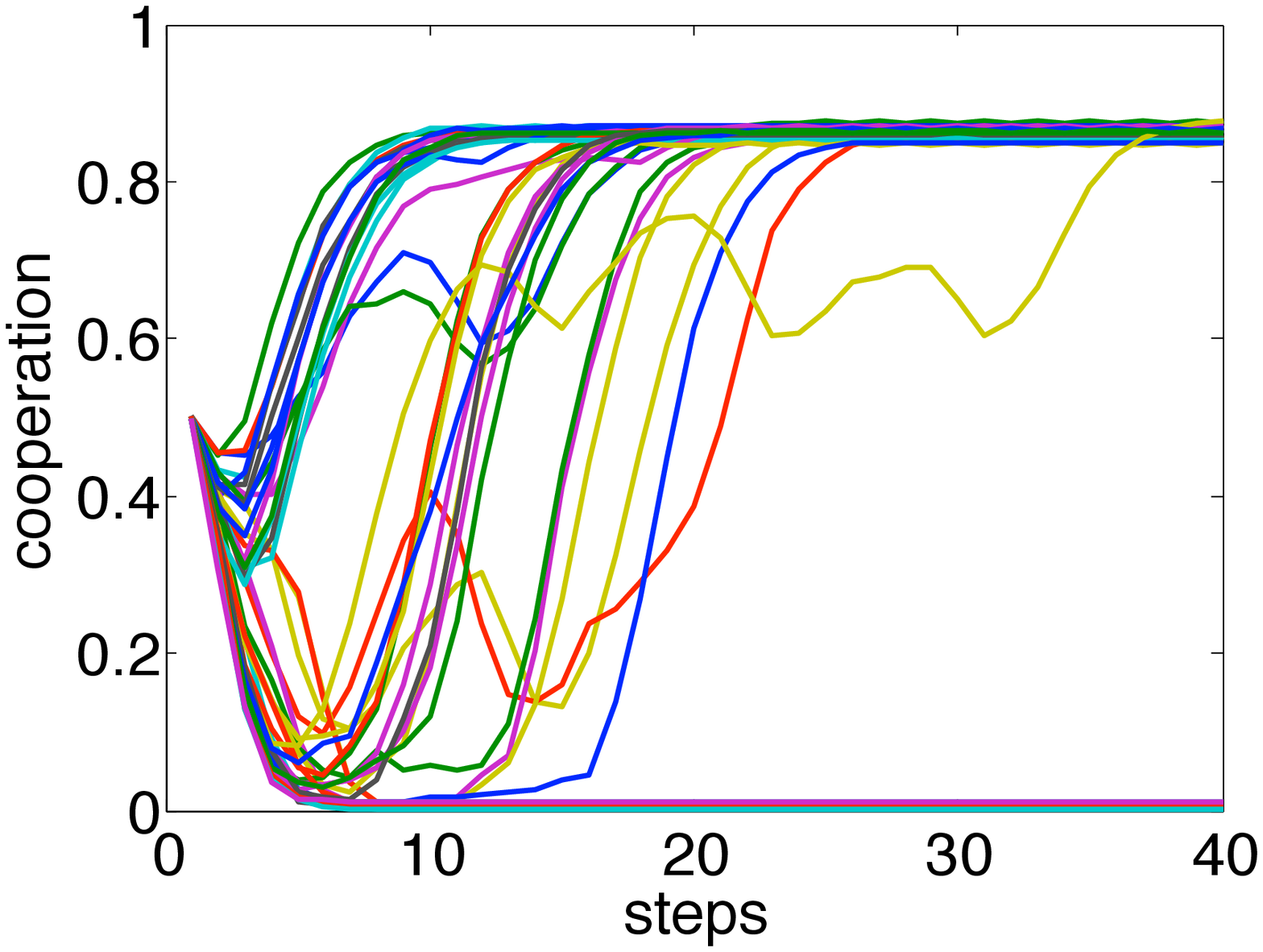}} \protect &
	\mbox{\includegraphics[width=5.5cm, height=4.2cm]{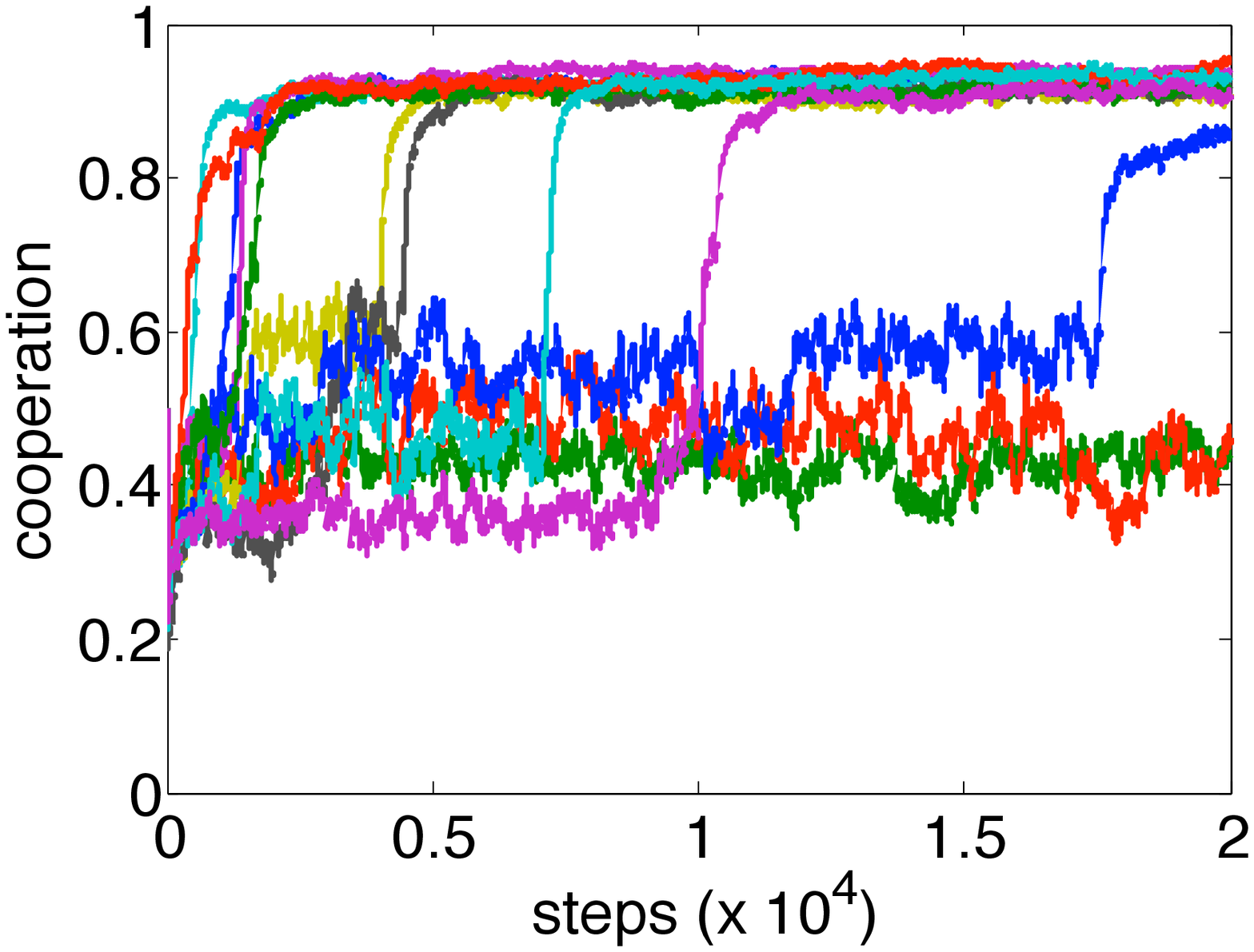}}   \protect \\
	(a)   &  (b) \\
	\multicolumn{2}{c}{\mbox{\includegraphics[width=5.5cm, height=4.2cm]{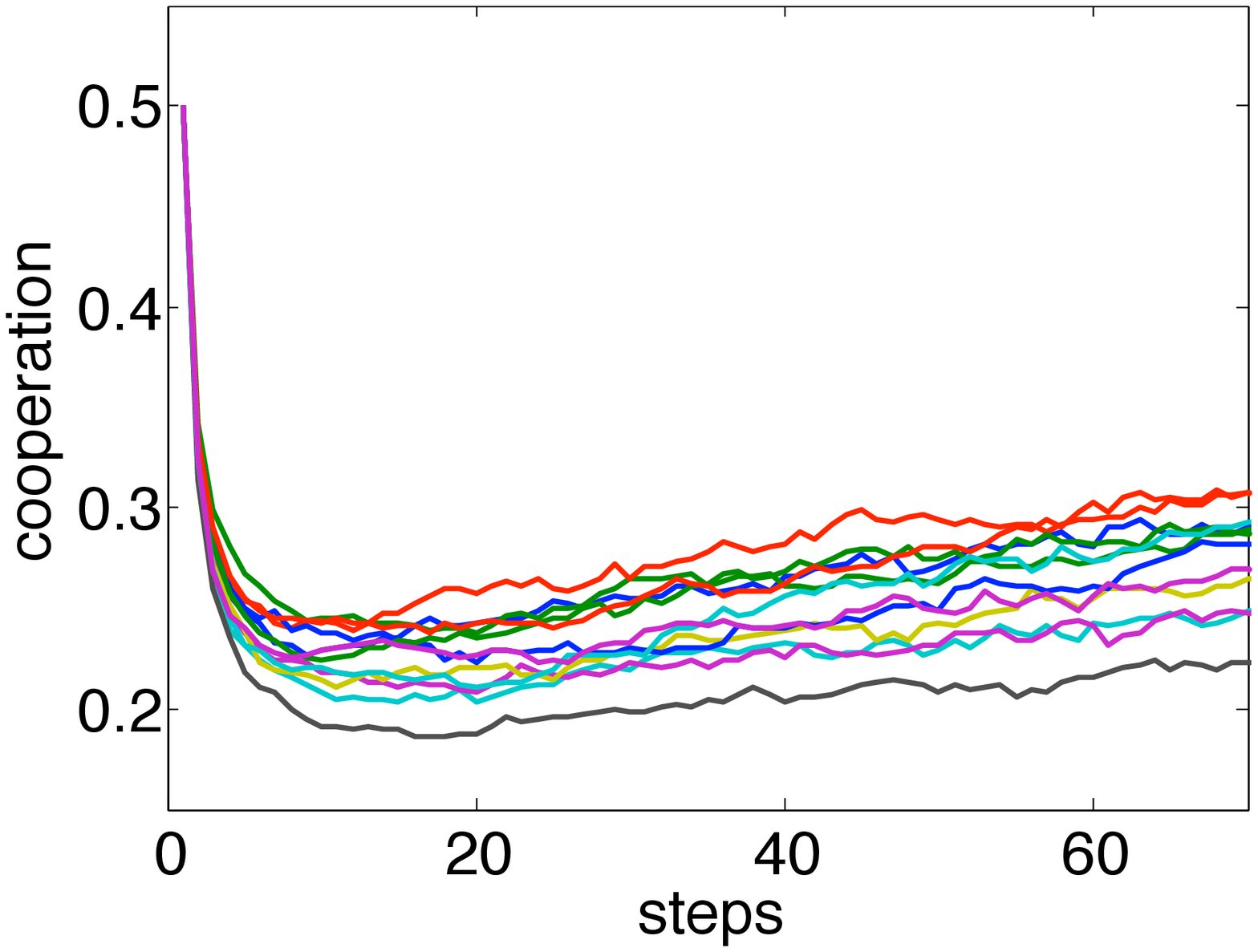}} \protect}\\
	\multicolumn{2}{c}{(c)} \\
\end{tabular}
\caption{PD time series with $b=1.8$; (a) imitation of the best; (b) replicator dynamics; (c) replicator dynamics (first 70 steps).\label{id-sep-runs}}
\end{center}
\end{figure}

Under discrete replicator dynamics rule with accumulated payoff, and using \textit{synchronous update}, Santos and Pacheco \cite{santos-pach-05} found that,
when compared to regular lattices, SF networks lead to high levels of cooperation for all values of
the parameters $b$ (for PD) and $r$ (for SD). These results have been reproduced by us and
are shown in the upper half of figure \ref{acc-payoff}. Cooperation is also much higher in SF
graphs than what has been obtained for Watts--Strogatz small-world
graphs \cite{social-pd-kup-01,tom-lut-gia-hd-06}. When using the ``imitation of the best''
strategy-switching rule with synchronous update and accumulated payoff the results are similar, as one can see in the lower part of figure \ref{acc-payoff},
although there is a marked fall in the high-b and high-r region with respect to replicator dynamics. However,
when one lingers on the standard deviations (represented as error bars in the figure), one
sees that the results for the imitate the best rule are noisy, with quite large fluctuations. Deviations are smaller for the replicator
dynamics, see figure \ref{acc-payoff}. The reason for the instability and the large fluctuations
 can be traced to the step function nature of the update
rule, as can be seen in figure \ref{id-sep-runs} (a), in which 
$40$ individual PD runs are plotted, all with $b=1.8$. In all runs cooperation falls at the beginning,
the cooperators then often recover but not always, as there are several runs (about $1/5$ for
the data used here) in which cooperation
never recovers. 
On the other hand, when using replicator dynamics, there is still a systematic drop of cooperation at the beginning (figure \ref{id-sep-runs} (c)),
nevertheless it tends to rise again in the long run, although this may happen
very late in the simulation (see figure \ref{id-sep-runs} (b)). To better observe this phenomenon, we have doubled the number of time steps ($2 \times 10^4$).

\begin{figure} [!ht]
\begin{center}
\begin{tabular}{cc}
	\mbox{\includegraphics[width=5.25cm, height=4cm]{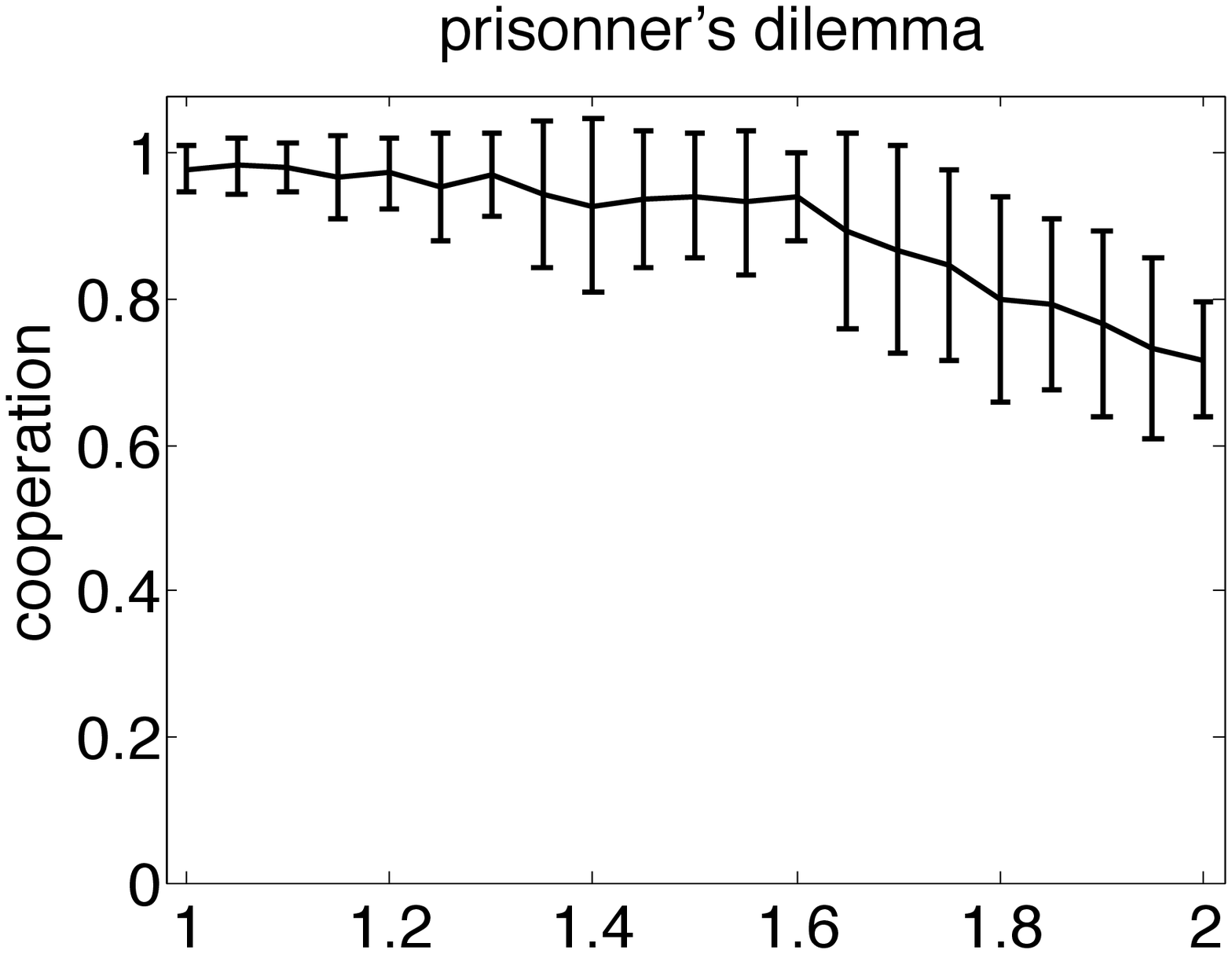}} \protect & 
	\mbox{\includegraphics[width=5cm, height=4cm]{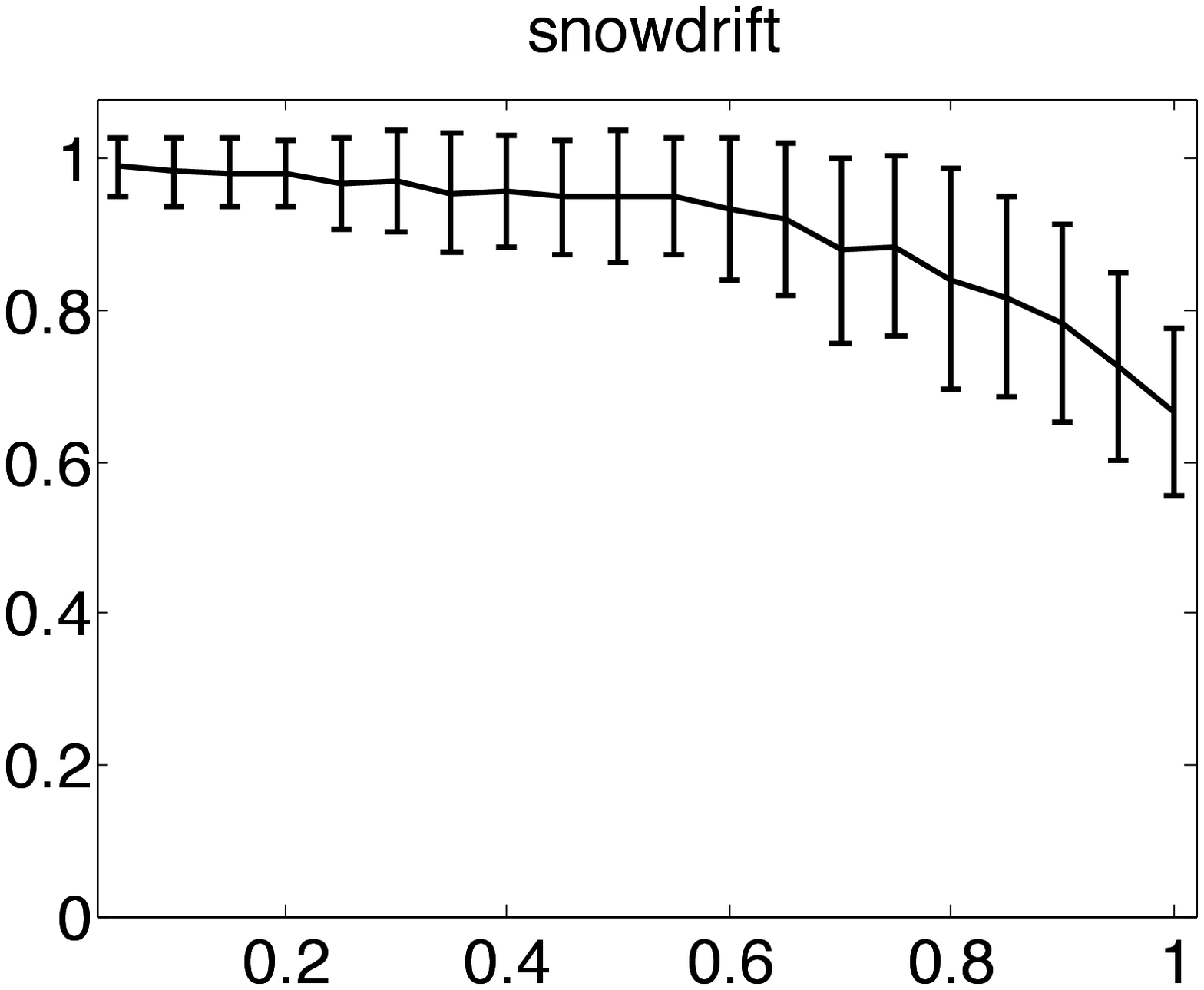}}   \protect \\
	\mbox{\includegraphics[width=5.25cm, height=3.9cm]{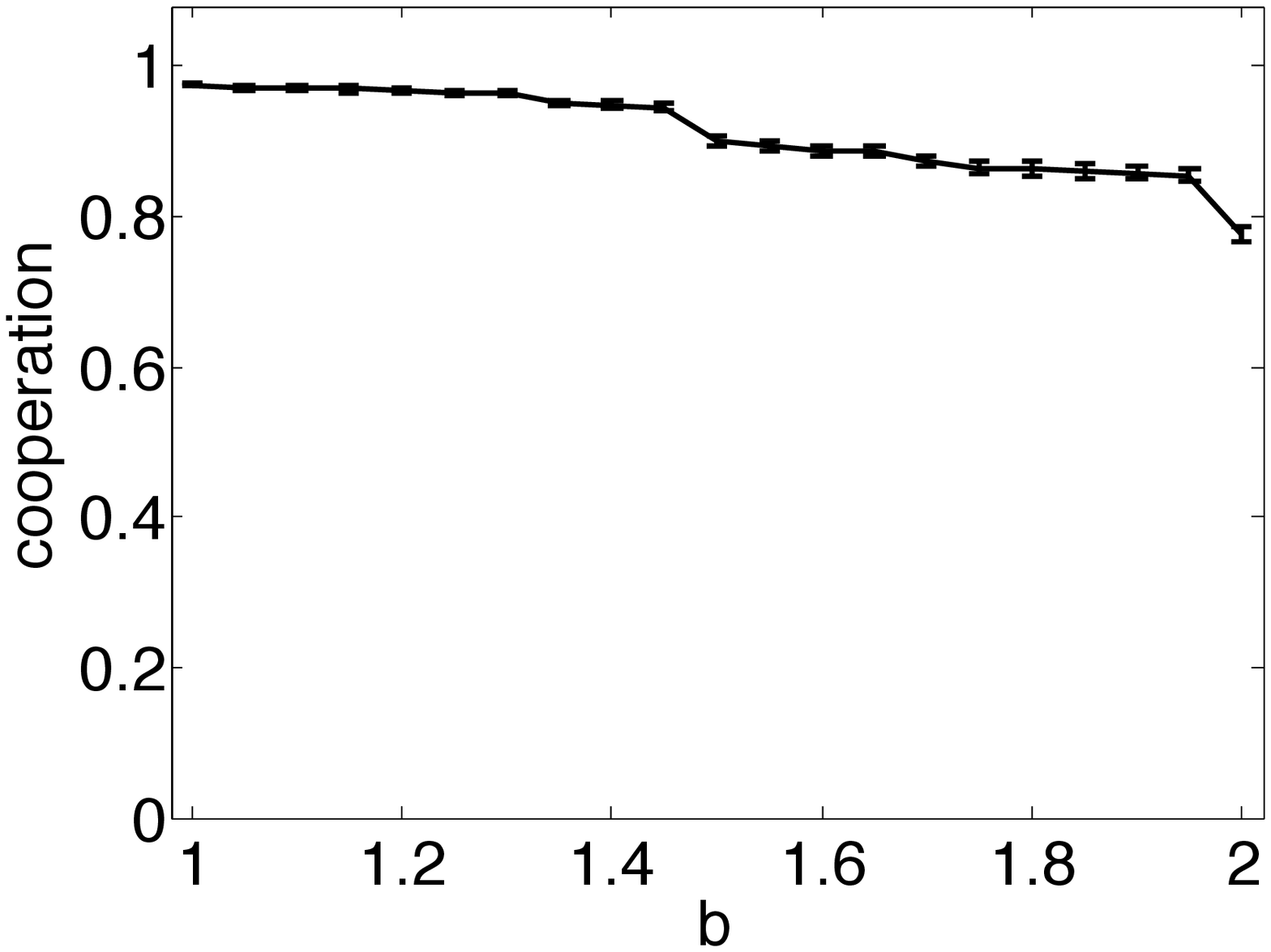}} \protect & 
	\mbox{\includegraphics[width=5cm, height=3.95cm]{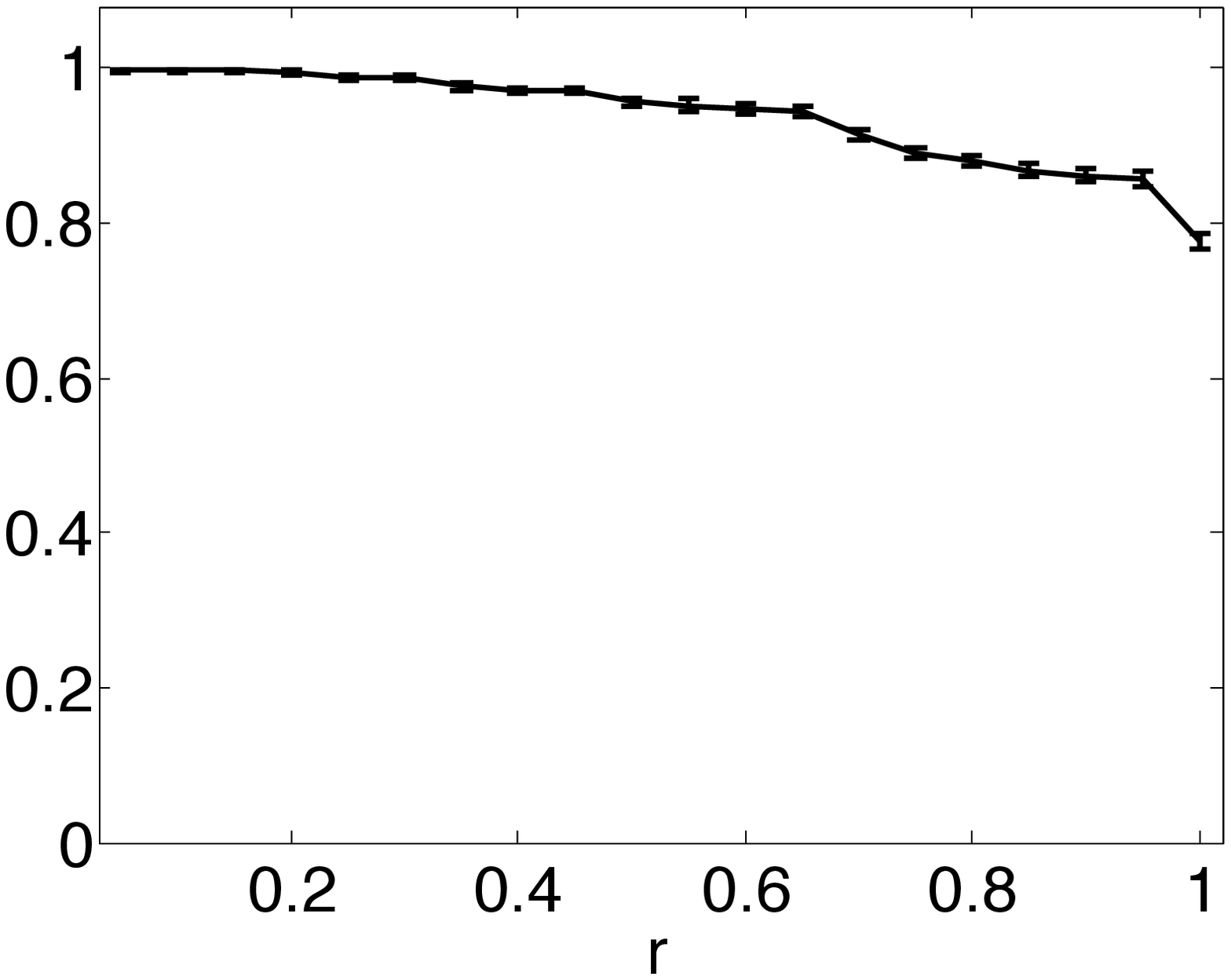}}   \protect \\
\end{tabular}
\caption{Fraction of cooperators on SF BA networks of size $10^4$ and average degree $\bar k=4$ with accumulated payoff and asynchronous dynamics. Mean values
over 50 runs. Upper figures: replicator dynamics; lower figures: imitation of the best. \label{async-acc-payoff}}
\end{center}
\end{figure}

We thus see that the results on BA SF graphs depend on the update rule,
although the level of cooperation is still higher than what is found on regular, Watts--Strogatz, and random graphs \cite{hauer-doeb-2004,tom-lut-gia-hd-06}.
However, we wish to point out that if we use an \textit{asynchronous  update} policy
\footnote{We use the standard uniform random choice (with replacement) of players in the
population, which is a discrete approximation of a Poisson process.} with the ``imitate the best'' rule instead
of the usual synchronous one, the result is a higher level of cooperation with far less
fluctuations than the synchronous case (compare lower parts of figures 
\ref{acc-payoff} and \ref{async-acc-payoff}).
 One might reason that the combination of
synchronous update and of ``imitate the best'' is fully deterministic, which implies that
particular chains of events, such as cascades of defection, will be amplified. Introducing
stochasticity through asynchrony in the update sequence strongly mitigates the likelihood of
such series of events.
On the other hand,
when using replicator dynamics, the lack of stochasticity in synchronous update is somehow
compensated for by the probabilistic strategy change rule, which could explain the similarity
of the results in this latter case (compare the upper parts of figures \ref{acc-payoff} and \ref{async-acc-payoff}
respectively). 

To illustrate the influence of timing when ``imitate the
best" is the rule used for strategy update, 
suppose that a defector occupies the most highly connected node in the graph and that it is surrounded
by cooperators exclusively. Then, at the next time step in synchronous update, all those
cooperators will turn into defectors. From there, a wave of defection could quickly
propagate through the network, leading to a state whereby cooperation cannot be recovered.
On the other hand, when players are updated in random order, only a fraction of the neighbors
will imitate the defector, at the same time lowering the payoff of the central defector,
and thus making it less attractive to be imitated in future encounters. This kind of process limits the propagation
of defection and allows cooperation to establish itself and be stable.
This highlights some
shortcomings of synchronous dynamics, which is unrealistic and may give rise to spurious
effects \cite{hubglance93}. Our conclusion is that, although there is often no 
significant difference between
synchronous and asynchronous update in evolutionary games, as it is the case here under
replicator dynamics, the latter is to be preferred
for reasons of generality and reliability. However, for the sake of comparison with previous
results, in the rest of the paper we use synchronous update.

\begin{figure} [!ht]
\begin{center}
\includegraphics[width=8.2cm]{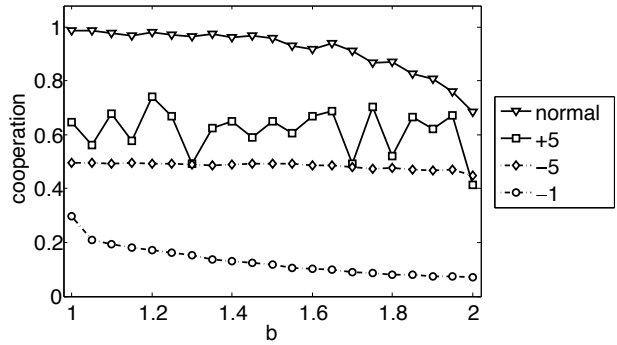}  
\caption{Fraction of cooperation for the PD game using replicator dynamics and accumulated payoff.
A translation of the payoff matrix can produce a fall in cooperation (shift of $-1$) as well as
unpredictable behaviors (shift of $+5$) with some runs containing high levels of cooperation and others ending up with massive defection.
Standard deviations are not plotted here to improve readability.\label{invariance}}
\end{center}
\end{figure}

Now we turn our attention to the assumption that a player's utility is the sum, i.e. the accumulated payoff of all
pair interactions with its nearest neighbors. Although this appears to be a logical step
to follow, we shall show that it may cause both conceptual and technical problems.
Obviously, one would assume that if an individual has more links to cooperators,
and that the payoffs are positive quantities,
she should earn more than another player with fewer cooperating neighbors.
However, this begs the question of how 
the network got there in the first place. BA SF graphs are incrementally built by using
linear preferential attachment \cite{alb-baraba-02}. In this model there is no cost
associated to the formation of a new link. However, although this model may be adequate for
citation networks or, to some extent, the Web, it is well known that this cannot be the
case in most other instances. Thus, other models have been proposed that take into account cost
and other factors in network formation \cite{newman-03}. In our case, it is as if
the population would be ``injected'' on an already full-grown, topology-favorable network,
while the rules of the game and other considerations necessarily should play a role in
the network formation and dynamics. The same remarks also hold for the ``configuration''
SF graphs, although these networks are built starting from the degree distribution and a fixed
number of nodes, rather than incrementally.
Furthermore, a technical problem arises when combining replicator dynamics with accumulated payoff.
In infinite mixing populations, classical evolutionary game theory states that replicator dynamics is invariant
under positive affine transformations of payoffs with merely a possible change of time scale \cite{weibull95}.
This invariance still holds in finite degree-homogenous populations.
However, when different individuals start having different degrees, things are not quite the same.
Let $\Pi_i$ denote a player $i$'s aggregated payoff. Furthermore, let $\phi(\Pi_j -\Pi_i) = (\Pi_j - \Pi_i)/(Dk_{>})$ be the
probability function according to which $i$ adopts neighbor $j$'s strategy,
with $D = max\{T, R, P, S\} - min\{T, R, P, S\}$ and $k_> = max\{k_i, k_j\},$ where $k_x$
represents the degree of \mbox{player $x$} \cite{santos-pach-05}.
If we now apply a positive affine transformation of the payoff matrix, this leads to the new  aggregated payoff
$\Pi_{i}' = \alpha \Pi_{i} + \beta k_i$
and hence $\phi(\Pi_{j}' - \Pi_{i}') = (\alpha\Pi_j + \beta k_j - \alpha\Pi_i - \beta k_i)/(\alpha Dk_>) = 
\phi(\Pi_{j} - \Pi_{i}) +   (k_j - k_i)/(\alpha Dk_>)$.
One can clearly see that using accumulated payoff does not lead to an invariance of the replicator dynamics under shifts of the payoff matrix. 
As an illustration of the violation of this invariance, figure \ref{invariance} shows cooperation curves 
for the PD when applying such payoff transformations.

This has several implications such as limiting the results obtained in 
\cite{santos-pach-05} strictly to the studied values of $b$ and $r$, and to an impossibility to rescale the payoff matrix. In a more recent study \cite{santos-pnas-06} Santos et al. investigated the same games
in a wider parameter space, but still using accumulated payoff, which again makes the results
non-invariant with respect to a positive affine transformation.
\begin{figure} [!ht]
\begin{center}
\begin{tabular}{cc}
	\mbox{\includegraphics[width=5.25cm, height=4cm]{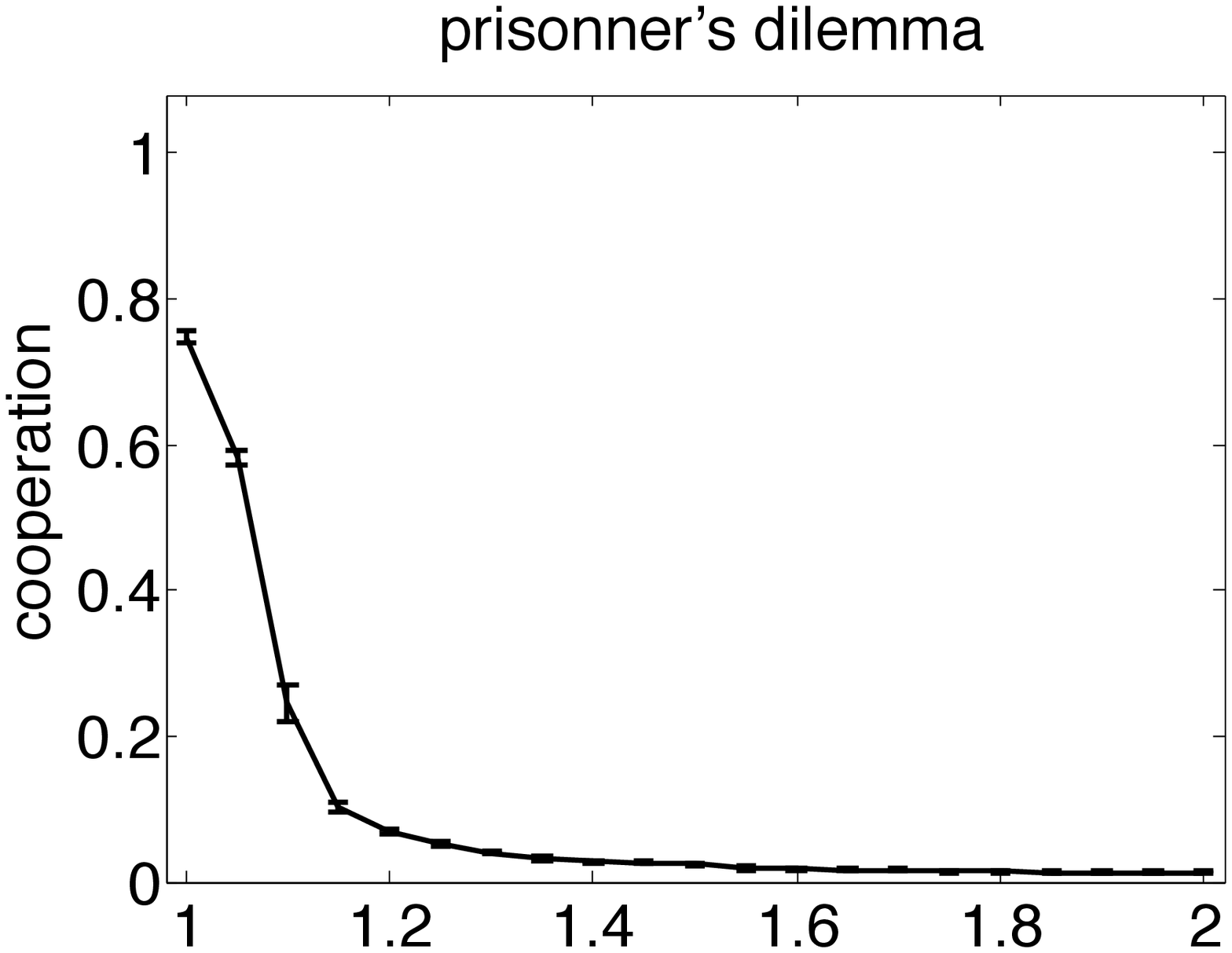}} \protect & 
	\mbox{\includegraphics[width=5cm, height=4.01cm]{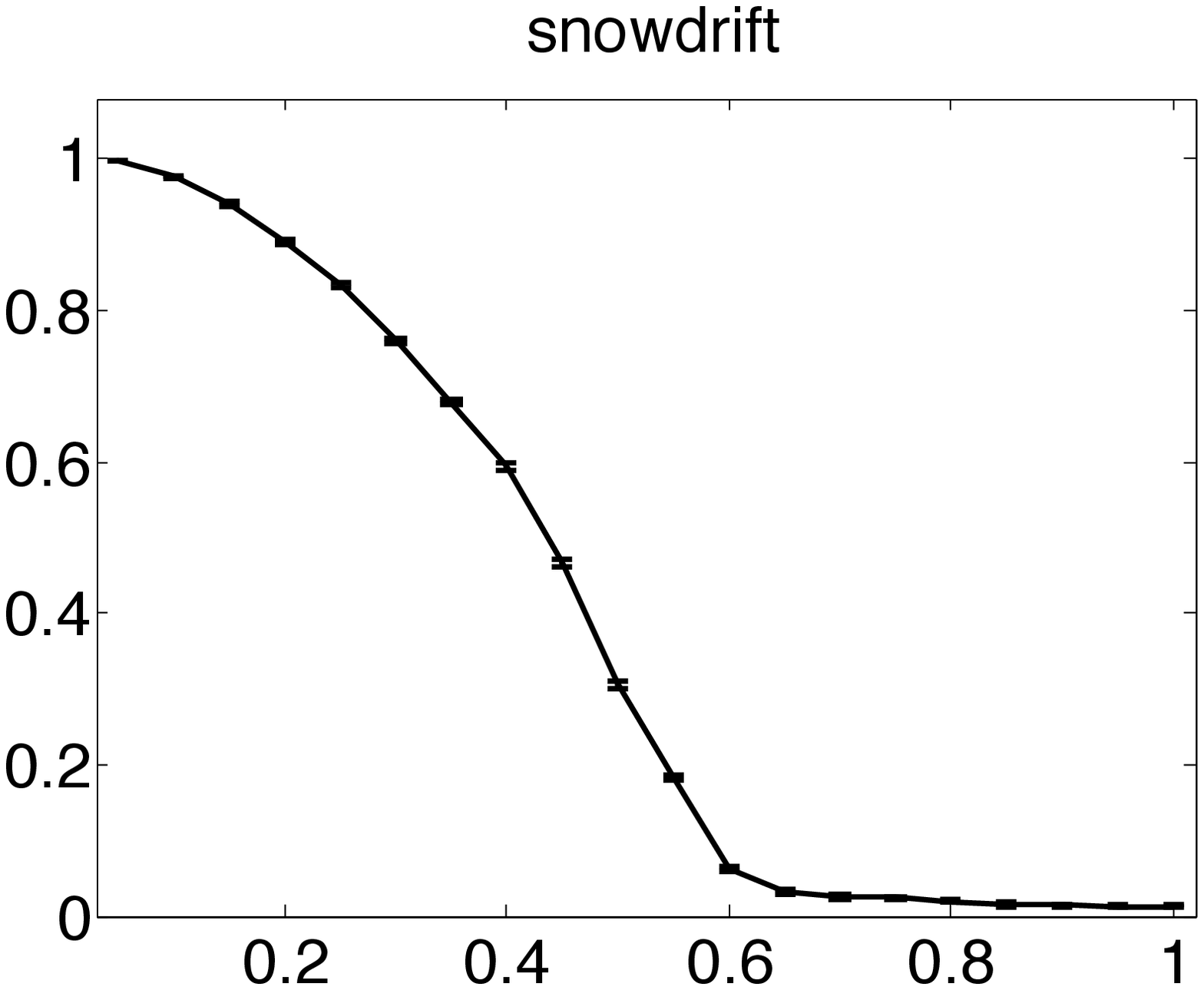}}   \protect \\
	\mbox{\includegraphics[width=5.25cm, height=3.9cm]{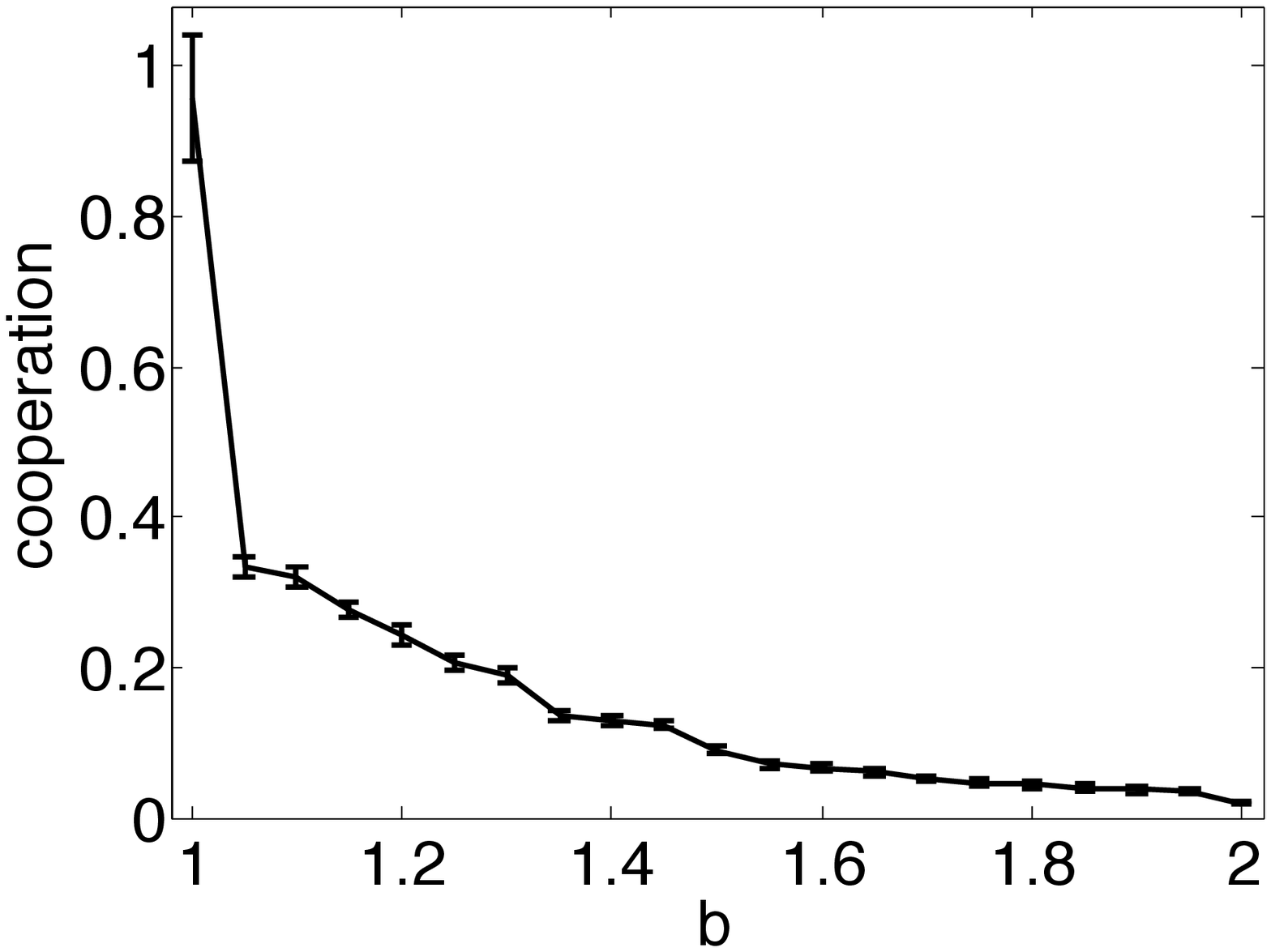}} \protect & 
	\mbox{\includegraphics[width=5.05cm, height=3.95cm]{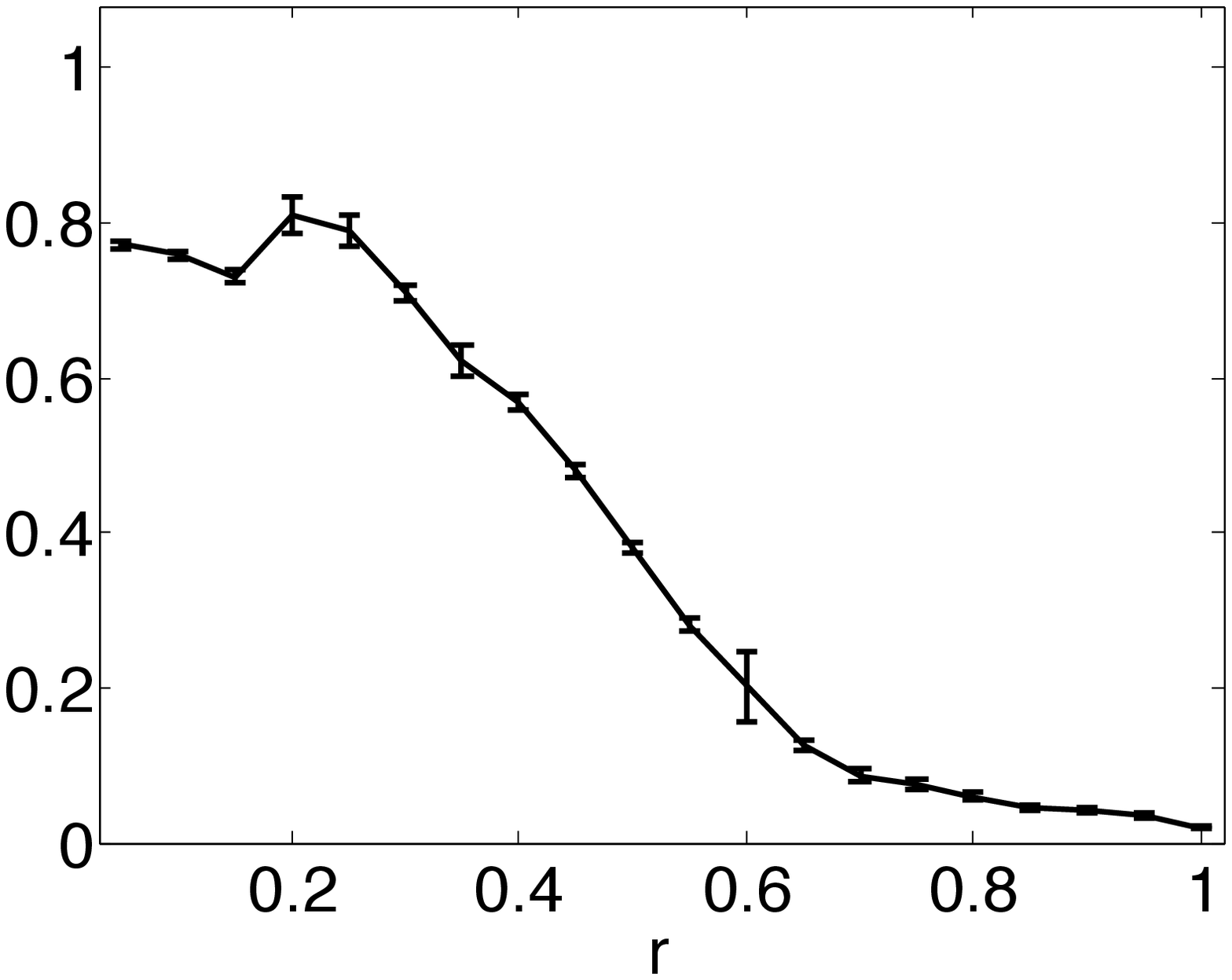}}   \protect \\
\end{tabular}
\caption{ Fraction of cooperators on SF BA networks of size $10^4$ with average degree $\bar k=4$ 
using average payoff and synchronous dynamics. Mean values
over 50 runs. Upper figures: replicator dynamics; lower figures: imitation of the best.\label{av-payoff}}
\end{center}
\end{figure}
Therefore, we
repeated the numerical simulations with \textit{average payoff}, i.e. the aggregated payoff
obtained by one player divided by the number of links the player has to nearest neighbors,
which, along with the shortcomings described above, has the advantage of
leaving the replicator dynamics invariant under positive affine transformations.

In figure \ref{av-payoff} we report results for the PD and SD games using average payoff with synchronous updating
dynamics, and the same parameter set as in \cite{santos-pach-05}.
Looking at the figures, and comparing them with the results of \cite{santos-pach-05}
(replicated here for $\bar k=4$ in figure \ref{acc-payoff}), one immediately sees that the
cooperation level reached after the transient equilibration period
is much lower, and comparable with the results found for regular and random graphs.
This is reasonable, given that now it is as if each individual had the same average number of
neighbors as far as its payoff is concerned.

\begin{figure} [!ht]
\begin{center}
\includegraphics[width=7.5cm]{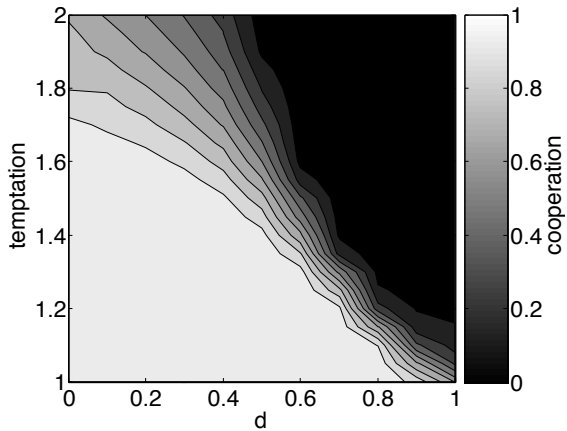} 

\caption{Cooperation level as a function of the parameter $d$ of equation \ref{transition_eq} in the PD for temptation values between
1 and 2. Cooperation prevails in light areas; Darker areas mean more defection. Results are
the average of 50 runs. \label{interp}}
\end{center}
\end{figure}

To reach a better understanding of the difference between accumulated and average payoff,
we interpolated between the two extreme cases according to the formula
\begin{equation}
\Pi_i = \frac{1}{k^d} \sum_{j} \pi_{i,j},
\label{transition_eq}
\end{equation}

\noindent where $d\in [0,1]$, $\Pi_i$ is the net payoff of player $i$, and
$\pi_{i,j}$ is the payoff player $i$ obtains when interacting with neighbor $j$. One 
can see that, when $d=0$ we recover the accumulated payoff
value, while $d=1$ corresponds to the average payoff case. Figure \ref{interp} clearly shows
that, as $d$ varies from 0 to 1, and thus the ratio varies from 1 to $1/k$, cooperation levels steadily decrease for all values of the temptation on the y-axis. So, the way in which individual
payoff is computed has a large influence on cooperation levels that can be reached, in the average,
on a given network topology. 

\section{Evolutionary Stability}
\label{stability}

\begin{figure} [!ht]
\begin{center}
\begin{tabular}{cc}
	\mbox{\includegraphics[height=4.5cm]{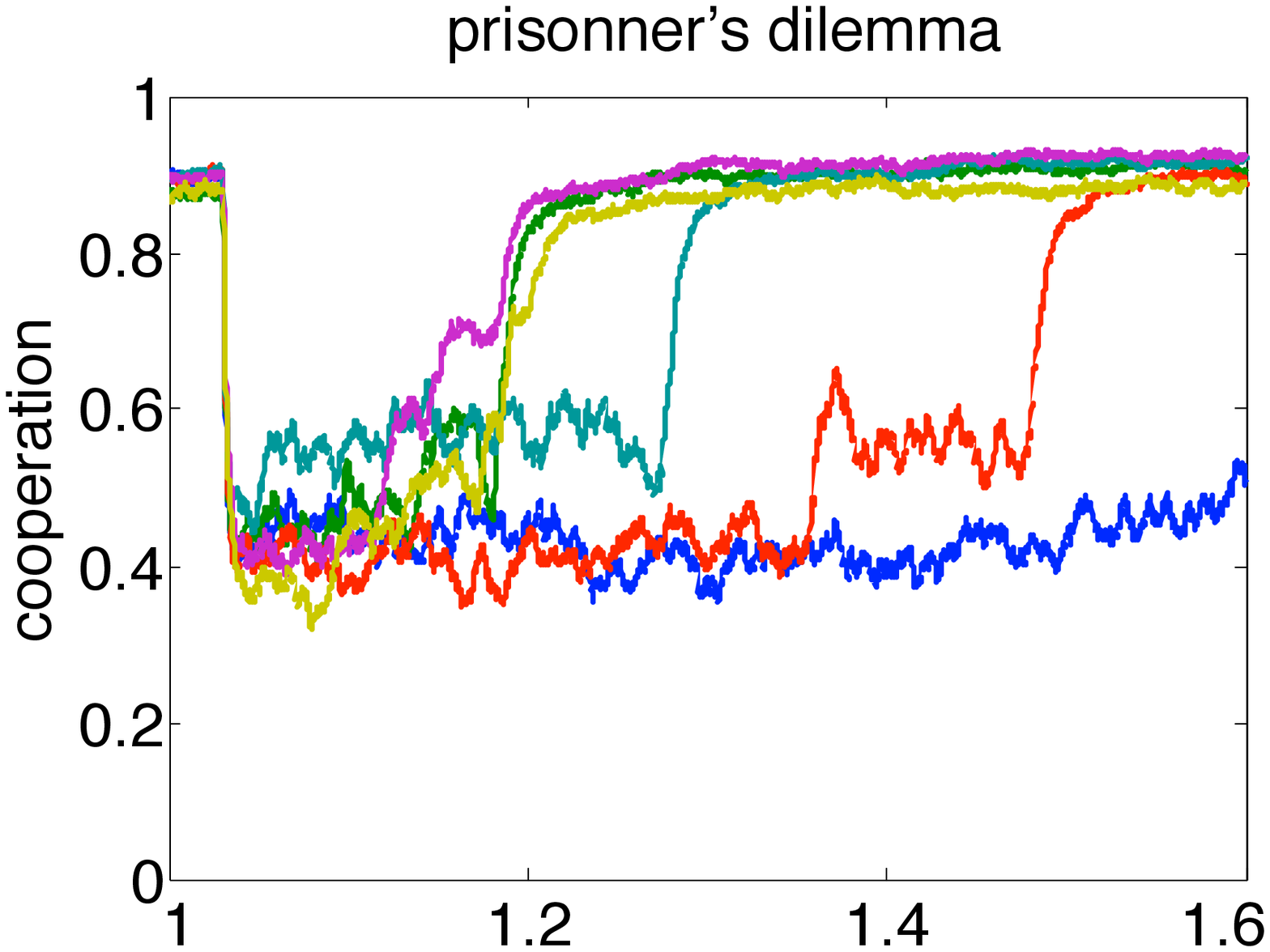}} \protect & 
	\mbox{\includegraphics[height=4.5cm]{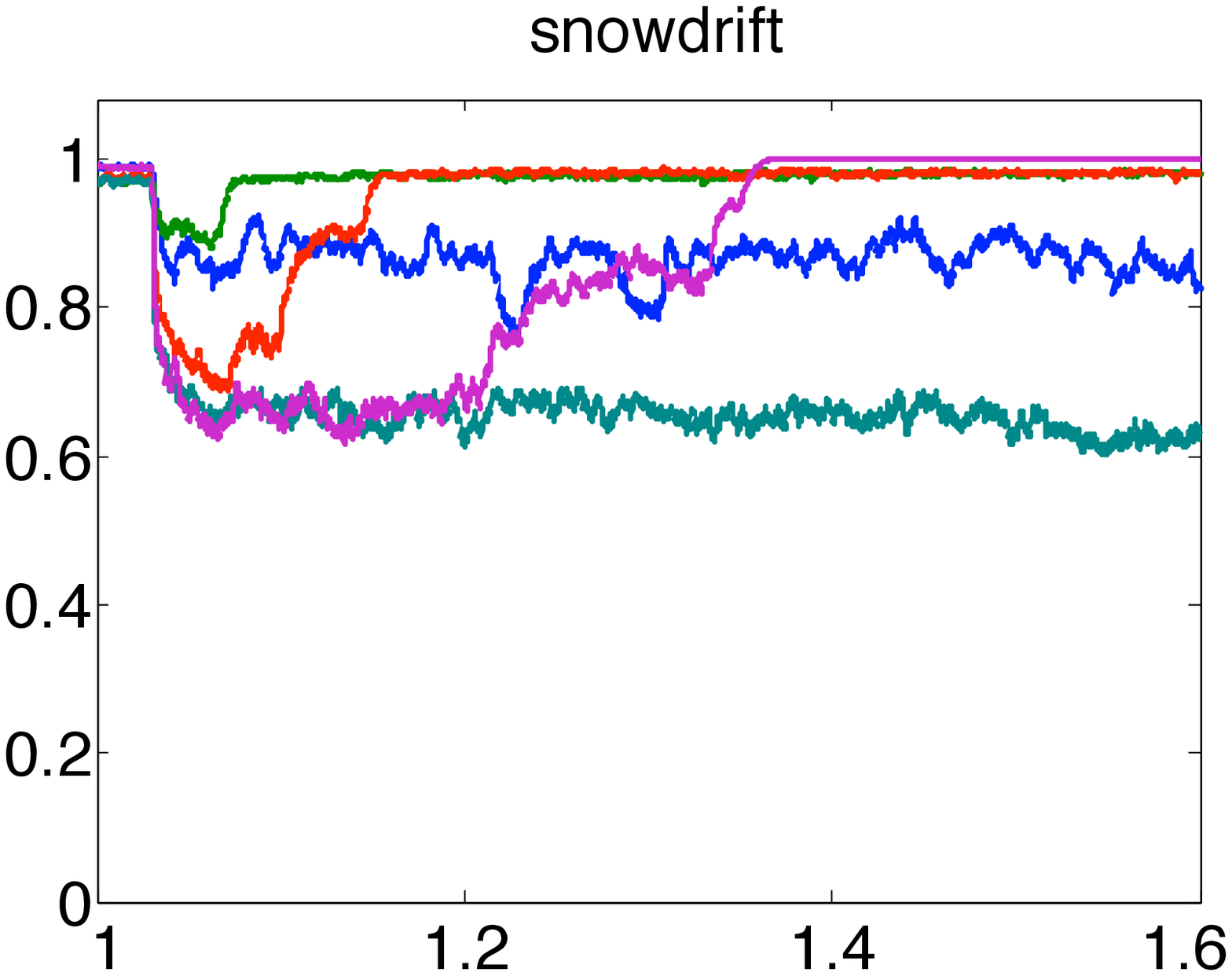}}   \protect \\
	\mbox{\includegraphics[height=4.5cm]{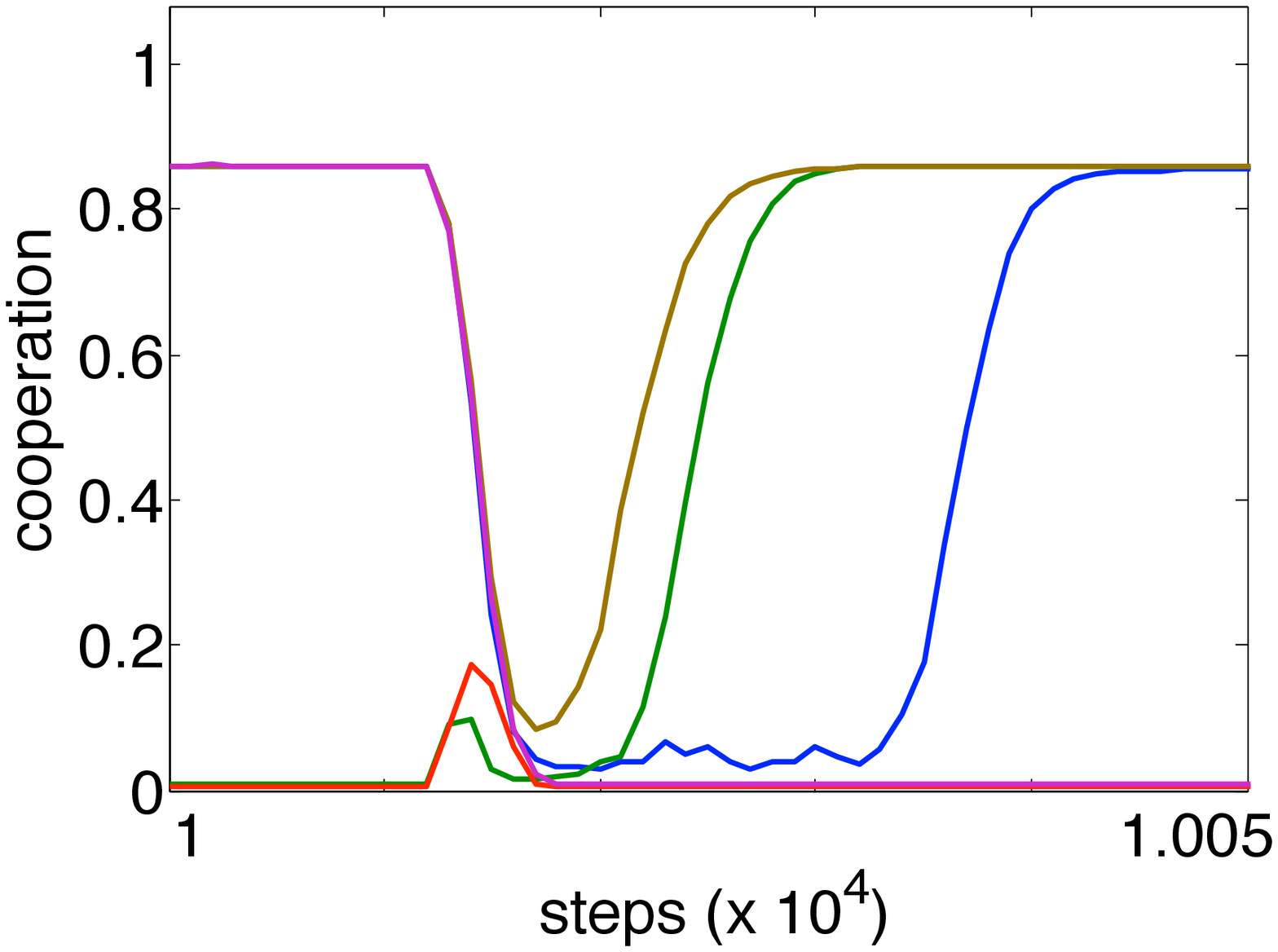}} \protect & 
	\mbox{\includegraphics[height=4.525cm]{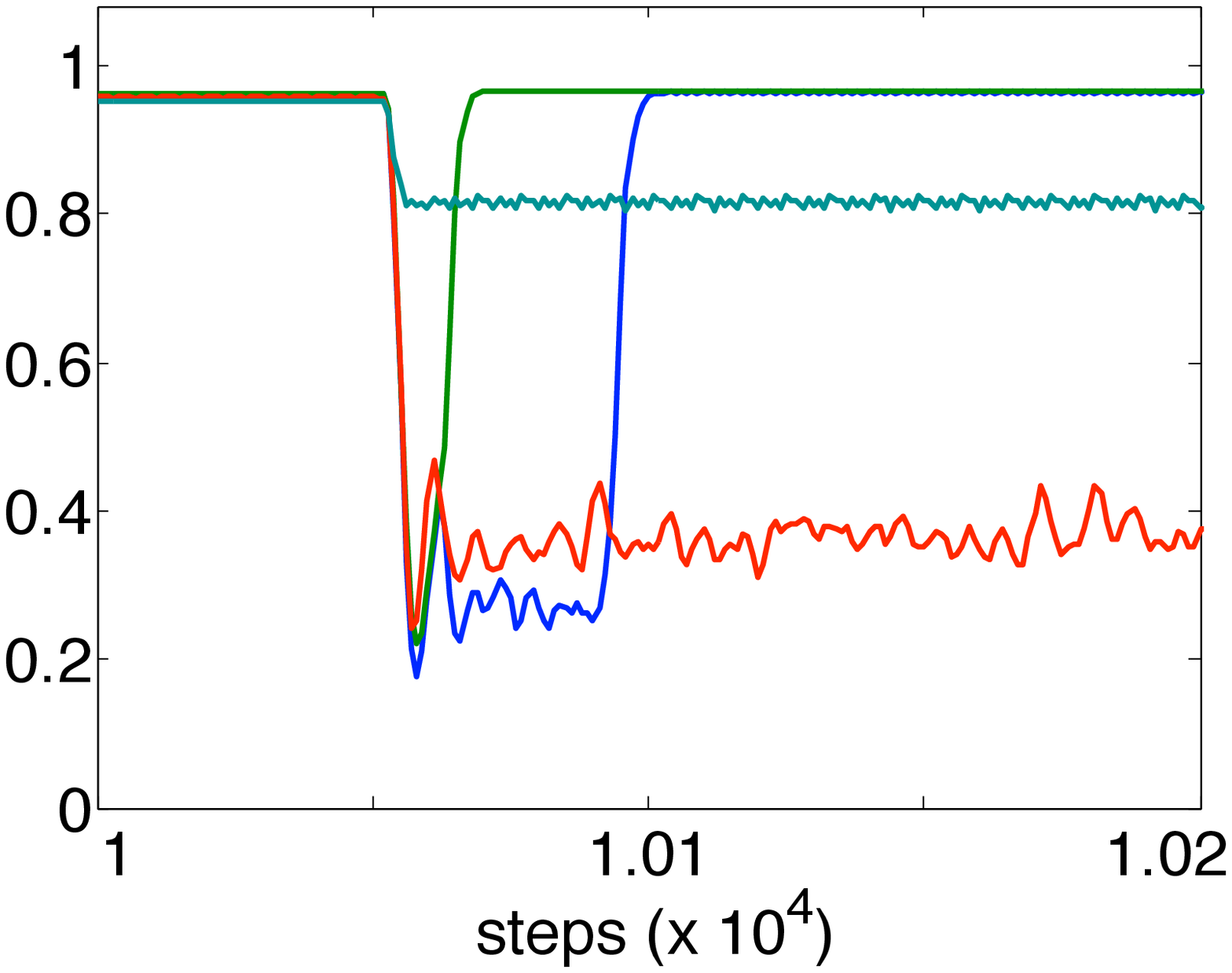}}   \protect \\
\end{tabular}
\caption{System stability when using accumulated payoff.
For each parameter set, 100 runs have been perturbed, but
only a few individual runs are plotted here to expose the behaviors encountered.
Upper figures: replicator dynamics; lower figures: imitation of the best.
Left-hand figures: $b=1.8$; right-hand figures: $r=0.5.$\label{ev-stab}}
\end{center}
\end{figure}

Evolutionary stability, i.e. the resistance to invasion by mutant strategies, is an important issue when dealing with evolutionary games \cite{weibull95}. The effect of switching the strategy
of the hub with largest connectivity in a totally cooperating population has been studied
in \cite{joeb-santos06}.
Here we use a different approach to perturb the population after it has reached a quasi-stable state by switching the strategy of a few players having the strategy of the greater number.
This was done
for values of $b \in \{1.2, 1.5, 1.8\}$ and $r \in \{0.2, 0.5, 0.8\}$.
We then give the system $6000$ time steps to attempt to reattain its initial stable state.
For reasons of space, we only plot the results obtained for $b=1.8$ and $r=0.5$ (see figure \ref{ev-stab}).
Given the scale-free nature of the interaction network, introducing a small amount of random noise does not have
any effect on the population stability. On the other hand, when cooperator hubs switch strategy (one to
five in our study), avalanches of defection can form and propagate through the population.
Under replicator dynamics and when using accumulated payoff, about $1/6$ of the PD runs do not recover the state
previously attained at time step $10^4$. This fraction rises to $1/3$ for the SD game.
With the imitation of the best rule, $1/10$ of the PD and SD runs
fail to recover from the perturbations.
In contrast to accumulated payoff, average payoff does not allow perturbations to generate any noticeable effect, i.e. the system remains quite stable.

\section{Conclusions}
\label{concl}

In conclusion, we have deepened and extended the study presented in \cite{santos-pach-05}
clarifying the role of the updating rule and the type of payoff attributed to
players. We have shown that the games are not invariant under linear affine tranformations
when using accumulated payoff, while average payoff does not have this problem, although
it may artificially reduce the impact of scale-free degree networks.
We have also seen that asynchronous update dynamics, being more likely in a system
of independently interacting agents, by eliminating artificial effects due to the
nature of synchronous update, may give rise to steadier quasi-equilibrium states.
Moreover, we have studied several dynamical aspects of the evolution of the populations
such as their transients before attaining the steady-state, and their evolutionary stability,
showing that scale-free networks of interactions provide a quite stable environment 
for the emergence of cooperation when using accumulated payoff, except when hubs are targeted by the mutations, in
which case a sizable number of runs do not recover the original state, at least within the
simulation times allowed in our numerical experiments.

\paragraph{Acknowledgments.} E. Pestelacci and M. Tomassini gratefully acknowledge financial support by the Swiss 
National Science Foundation under contract 200021-111816/1.

{\small
\bibliographystyle{unsrt}
\bibliography{games}
}

\end{document}